\documentclass[12pt]{iopart}
\usepackage{iopams}  
\usepackage{graphicx,epsfig}
\usepackage{hyperref}
\usepackage{color}
\usepackage{textcomp}
\usepackage{upquote}
\renewcommand \thesection{\Roman{section}}
\renewcommand \thesubsection{\Alph{subsection}}
\newcommand{\beq}{\begin{equation}}
\newcommand{\eeq}{\end{equation}}
\newcommand{\ec}{\end{center}}
\newcommand{\bc}{\begin{center}}
\newcommand{\eea}{\end{eqnarray}}
\newcommand{\bea}{\begin{eqnarray}}
\newcommand{\eeas}{\end{eqnarray*}}
\newcommand{\beas}{\begin{eqnarray*}}
\newcommand{\bd}{\begin{description}}
\newcommand{\ed}{\end{description}}
\newcommand{\vx}{\mathbf{x}}

\newcommand{\mb}{\mathbf}
\newcommand{\vk}{\mathbf{k}}
\newcommand{\vko}{\mathbf{k}_1}
\newcommand{\vkt}{\mathbf{k}_2}

\newcommand{\vqo}{\mathbf{q}_1}
\newcommand{\vkth}{\mathbf{k}_3}
\newcommand{\vkf}{\mathbf{k}_4}

\newcommand{\vqf}{\mathbf{q}_4}

\newcommand{\vq}{\mathbf{q}}

\newcommand{\mysubsection}[2][]

\begin{document}
\title{Hyperskewness of $(1+1)$-dimensional KPZ Height Fluctuations}
\author{Tapas Singha \& Malay K. Nandy}
\address{Department of Physics, Indian Institute of Technology Guwahati, Guwahati 781039, India.}
\ead{s.tapas@iitg.ernet.in \& mknandy@iitg.ernet.in}
\vspace{10pt}
\date{(November, 12 2015)}

\begin{abstract}
We evaluate the fifth order normalized cumulant, known as hyperskewness, of height fluctuations dictated by the
$(1+1)$-dimensional KPZ equation for the stochastic growth of a surface on a flat geometry in the stationary state. 
We follow a diagrammatic approach and invoke a renormalization scheme to calculate the fifth cumulant given by a 
connected loop diagram. This, together with the result for the second cumulant, leads to the hyperskewness value
$\widetilde{S}=0.0835$. 
\end{abstract}
\hspace{4pc}

\section{Introduction}
The dynamics of surface growth has extensively been studied in nonequilibrium statistical 
physics in the last few decades \cite{book_stanley,krug97,Halpin95,Family_Physica}. To describe 
a local surface growth, Kardar, Parisi and Zhang \cite{KPZ86} (KPZ) first proposed a prototypical nonlinear equation 
for the fluctuating height field $h(\mb x, t)$, expressed as 
\beq
\frac{\partial } {\partial t} h(\mb x, t) =\nu_0
\nabla^{2} h+\frac{\lambda_0}{2} (\nabla h)^{2}+ \eta(\mb x, t),
\label{eq-kpz}
\eeq
where $\nu_0$ is the surface tension that smoothens the surface curvature and minimizes the surface area and $\lambda_0$
is a coupling constant representing the strength of the non-linear term. The stochastic term $\eta(\vx,t)$ is modeled as a
Gaussian white noise of zero average, $\langle \eta(\vx,t) \rangle=0$, and its covariance  
\beq
\langle \eta(\vx,t) \eta(\vx',t')\rangle=2 D_0 \delta^d(\vx-\vx')\delta(t-t')
\eeq
where $d$ is the substrate dimension. We shall consider the stationary state for the growth of a surface on a flat substrate of 
dimension $d=1$.

There are various systems that are mathematically equivalent to the KPZ dynamics. A few examples are: vorticity free noisy 
Burgers equation \cite{forster_pra_16_732_77}, stochastic heat equation (SHE), diffusion equation governed by random source and sink, 
directed polymer in random potentials \cite{J-krug_92}, in random media \cite{kpz87,Fisher_PRB_43_10728}, the sequence alignment 
of gene or protein \cite{Hwa_Lassig_PRL_76_2591,Hwa_gene_allignment_nature_399} etc. Various growth phenomena belong to 
the $(1+1)$-dimensional KPZ universality class (on the basis of scaling exponents \cite{book_stanley}). For example, 
growth of a bacteria colony \cite{Vicsek_90,Huergo_10}, fluid flow in a porous media \cite{Rubio_89}, 
turbulent liquid crystal \cite{K_A_Takeuchi,kazumasa}, slow combustion of a sheet of paper 
\cite{Myllys_PRE.64.036101,miettinena_epjb_46_55_05} etc. 

The $(1+1)$-dimensional KPZ equation satisfies the fluctuation dissipation theorem. The standard RG treatment of the $(1+1)$-dimensional 
KPZ equation leads to the same scaling exponents for the renormalized noise amplitude $D(k)$ and the renormalized surface tension $\nu(k)$.  
The resulting roughness and  dynamic exponents are $\chi=\frac{1}{2}$ and $z=\frac{3}{2}$ respectively \cite{KPZ86,kpz87} that satisfies the
scaling relation $\chi+z=2$. There have been various numerical models, namely, ballistic deposition (BD)\cite{Meakin86,Family_Physica}, 
Eden model \cite{M_Eden,plischke_prl_84,jullien_85,eden_plischke_85}, restricted solid on solid model (RSOS)\cite{Meakin93}, 
single step model (SSM) \cite{Meakin86,Plischek_87}, polynuclear growth (PNG) \cite{saarloos_86,goldenfeld,krug_pra_88,Michael_Herbert_00}
which have the same scaling exponents, and consequently, belong to the universality class of the $(1+1)$-dimensional KPZ equation. 

The identification of the universality class of experimental processes and numerical models have long been  based on the numerical values of 
the scaling exponents. New insights have been gained by Pr{\"a}hofer and Spohn \cite{Michael_Herbert_00} via the study of PNG model by mapping 
the problem to the longest permutation of random  Gaussian matrices \cite{book_Baik-Rains}. Thus they \cite{Michael_Herbert_00} 
estimated  different probability distributions which are close to Gaussian unitary ensemble 
(GUE) Tracy-Widom (TW), Gaussian orthogonal ensemble (GOE) TW, and Baik-Rains $F_0$ distribution for the curved, flat and random initial conditions,
respectively. The effects of these initial conditions on the growth have further been studied in various works. GUE TW, $F_2$, distribution has been 
observed for sharp-wedge initial condition \cite{sasamoto_prl_104_230602_10,2010JSMTE_11_013S}. Calabrese and Doussal \cite{calabrese_prl_106_250603_11}
mapped the one end free directed polymer to the flat initial condition KPZ equation and obtained GOE TW, $F_1$ distribution. A universal crossover 
function from GOE TW $F_1$ to Baik-Rains $F_0$ distribution has been studied in the experiment of turbulent liquid crystal (TLC) and numerical simulation of the 
PNG model by Takeuchi \cite{Takeuchi_prl_110_2013}. Imamura and Sasamoto \cite{Imamura_prl_108_2012} considered a two-sided Brownian motion as an 
 initial condition and obtained an exact solution of the $(1+1)$-dimensional KPZ equation as a function of time $t$ which goes to the 
 Baik-Rains distribution asymptotically. Halpin-Healy and Lin \cite{Healy_Lin_PRE.89.010103} have considered the numerical models 
 (RSOS, BD, SHE with multiplicative noise, SSM) belonging to the $(1+1)$-dimensional KPZ universality class and studied the statistics in the stationary
 state corresponding to the $(1+1)$-dimensional KPZ height fluctuations. The estimated stationary state distribution functions from these  
 numerical models yield a close comparison to the Baik-Rains $F_0$ distribution. Hence, they established that the stationary state $(1+1)$-dimensional
 KPZ-type height fluctuations obey the universal Baik-Rains $F_0$ distribution. Halpin-Healy and Takeuchi \cite{Healy-Takeuchi15} 
 performed Euler numerical integration  of $(1+1)$-dimensional KPZ equation considering a large system size (for flat $L=250,000$ and 
 stationary $L=10^4$) and large number of statistical realizations (for flat $4000$, curved $25,000$ and stationary $10^5$). They obtained distributions that agree well with the TW GOE, TW-GUE and Baik-Rains $F_0$ distributions for flat, curved and 
stationary initial conditions, respectively. Thus the probability distributions of the $(1+1)$-dimensional KPZ height fluctuations is governed by 
 the nature of the initial conditions despite having the same scaling exponents \cite{PhysRevE.64.036110}. In principle the probability distribution 
function can be obtained from the solution of the Fokker-Planck version \cite{huse_85,Parisi_1990} of the KPZ equation which is next to impossible
due to the nonlinear term. A more viable way to obtain a partial information of the PDF is to calculate the higher order cumulants such as third, fourth
and fifth cumulants. These cumulants, when normalized with respect to the second cumulant, yield skewness, kurtosis and hyperskewness, respectively. 

In this paper, we employ a diagrammatic approach and an  RG scheme to calculate the fifth cumulant. We follow an RG approach without rescaling 
that was found to be successful for calculating the skewness \cite{Singha-Nandy14_PRE.90.062402} and kurtosis \cite{Singha-Nandy15_jstatmech}. 
We evaluate the loop diagram for the fifth cumulant in the large scale and long time limits that yields $\tilde S=0.0835$. 

The rest of the paper is organized as follows. In Section II, the definition of hyperskewness in terms of moments and cumulants are presented. 
In Section III the calculations of fifth cumulant and the resulting hyperskewness is presented. Finally, a Discussion and Conclusion appear in  
Section-IV. 
\section{Moments and Cumulants}
Relations between moments and cumulants can be obtained by means of generating functions. 
The moment generating function $Z(\beta)$ and cumulant generating function $F(\beta)$ are defined as 
\beq
Z(\beta) \equiv \langle e^{\beta h} \rangle=\sum^{\infty}_{n=0}
\frac{\langle h^n\rangle}{n!} \beta^n
\label{gene-func}
\eeq
\beq
F(\beta) =\ln Z(\beta) =\sum^{\infty}_{n=1} \frac{\langle h^n\rangle_c}{n!} \beta^n,
\eeq
respectively where $\langle h^n \rangle$ is the $n$th moment and $\langle  h^n  \rangle_c$ is the $n$th cumulant, 
the angular bracket denotes an average with respect to the probability distribution. 
The relations among the first few moments and cumulants are expressed as
\begin{eqnarray}
\langle h \rangle=&&\langle h \rangle_c  \nonumber \\
\langle h^2 \rangle=&&\langle h^2 \rangle_c + \langle h \rangle^2_c \nonumber
\\
\langle h^3 \rangle=&&\langle h^3 \rangle_c +3 \langle h \rangle_c\langle h^2
\rangle_c +\langle h \rangle^3_c \nonumber \\
\langle h^4 \rangle=&&\langle h^4 \rangle_c +4\langle h \rangle_c \langle h^3
\rangle_c+3 \langle h^2 \rangle^2_c +6\langle h \rangle^2_c \langle h^2
\rangle_c + \langle h \rangle^4_c  \nonumber \\
\langle h^5 \rangle=&&\langle h^5 \rangle_c + 5\langle h \rangle_c
\langle h^4 \rangle_c +10 \langle h^2 \rangle_c \langle h^3 \rangle_c +10
\langle h\rangle^2_c \langle h^3 \rangle_c \nonumber \\
&&+15 \langle h\rangle_c \langle h^2 \rangle^2_c+10 \langle h\rangle^3_c \langle h^2 \rangle_c+\langle h \rangle^5_c.
\end{eqnarray}
 Since $h$ represents height fluctuations with respect to the mean height, $\langle h(\vx,t) \rangle=0$ in the stationary state.
Consequently, the relevant quantities for hyperskewness are 
\beq
\langle h^2 \rangle_c=\langle h^2 \rangle
\eeq
and
\beq
\langle h^5 \rangle_c= \langle h^5 \rangle-10 \langle h^2 \rangle \langle
h^3 \rangle.
\eeq
Hyperskewness $\widetilde{S}$ is defined as 
\beq
\widetilde{S}=\frac{\langle h^5\rangle_c
}{\langle h^2\rangle_c^{5/2}}=\frac{\langle h^5\rangle }{\langle
h^2\rangle^{5/2}}-10
\frac{\langle h^3\rangle }{\langle h^2\rangle^{3/2}}.
\label{defihypersk}
\eeq
Contribution to $n$th cumulant $\langle h^n(\vx,t) \rangle_c$ comes from  connected loop diagrams 
with $n$ external legs. 
\section{The Fifth Cumulant}
In this section, we calculate the fifth order cumulant via a renormalization scheme at one loop order. 
The Fourier transformation of $h(\vx, t)$ is expressed as
\beq
h(\vx,t)=\int \frac{d^{d}k \ d\omega}{(2\pi)^{d+1}} h(\vk,\omega) e^{i(\vk \cdot \vx-\omega t)}
\eeq
The Fourier transformation of Eq. \ref{eq-kpz} leads to obtain the following form as 
\begin{equation}
(-i \omega+\nu_0 k^2) h(\vk,\omega)=\eta(\vk,\omega)
-\frac{\lambda_0}{2}
\int \frac{d^d\vq
d\Omega}{(2\pi)^{d+1}}[\vq\cdot(\vk-\vq)]h(\vq,\Omega)h(\vk-\vq,\omega-\Omega)
\label{KPZFT}
\end{equation}
We shall treat $[-i \omega +\nu_0 k^2]^{-1}=G_0(k,\omega)$ as the bare propagator.
The Fourier transformation of fifth cumulant is expressed  as 
\begin{eqnarray}
\langle h^5(\vx,t)\rangle_c=&& \int \frac{d^{d+1} \hat{k}_1}{(2\pi)^{d+1}}\int
\frac{d^{d+1} \hat{k}_2}{(2\pi)^{d+1}} \!\int
\frac{d^{d+1} \hat{k}_3}{(2\pi)^{d+1}} \!\int \frac{d^{d+1}
\hat{k}_4}{(2\pi)^{d+1}} \!\int
\frac{d^{d+1} \hat{k}_5}{(2\pi)^{d+1}} \nonumber  \\
&& e^{i(\hat{k}_1+\hat{k}_2+\hat{k}_3+\hat{k}_4+\hat{k}_5)\cdot \hat{x}} \ 
\langle h(\hat{k}_1) h(\hat{k}_2) h(\hat{k}_3) h(\hat{k}_4) h(\hat{k}_5)
\rangle_c
\label{h5c}
\end{eqnarray}
where $\hat{x}\equiv (\mb x,t)$ and $\hat{k}_i\equiv (\mb k_i,\omega_i).$ The connected loop diagram corresponding to the fifth order 
cumulant is shown in Fig.\ 1.

\begin{figure}[h!t]
\begin{center}
\includegraphics[width=5cm]{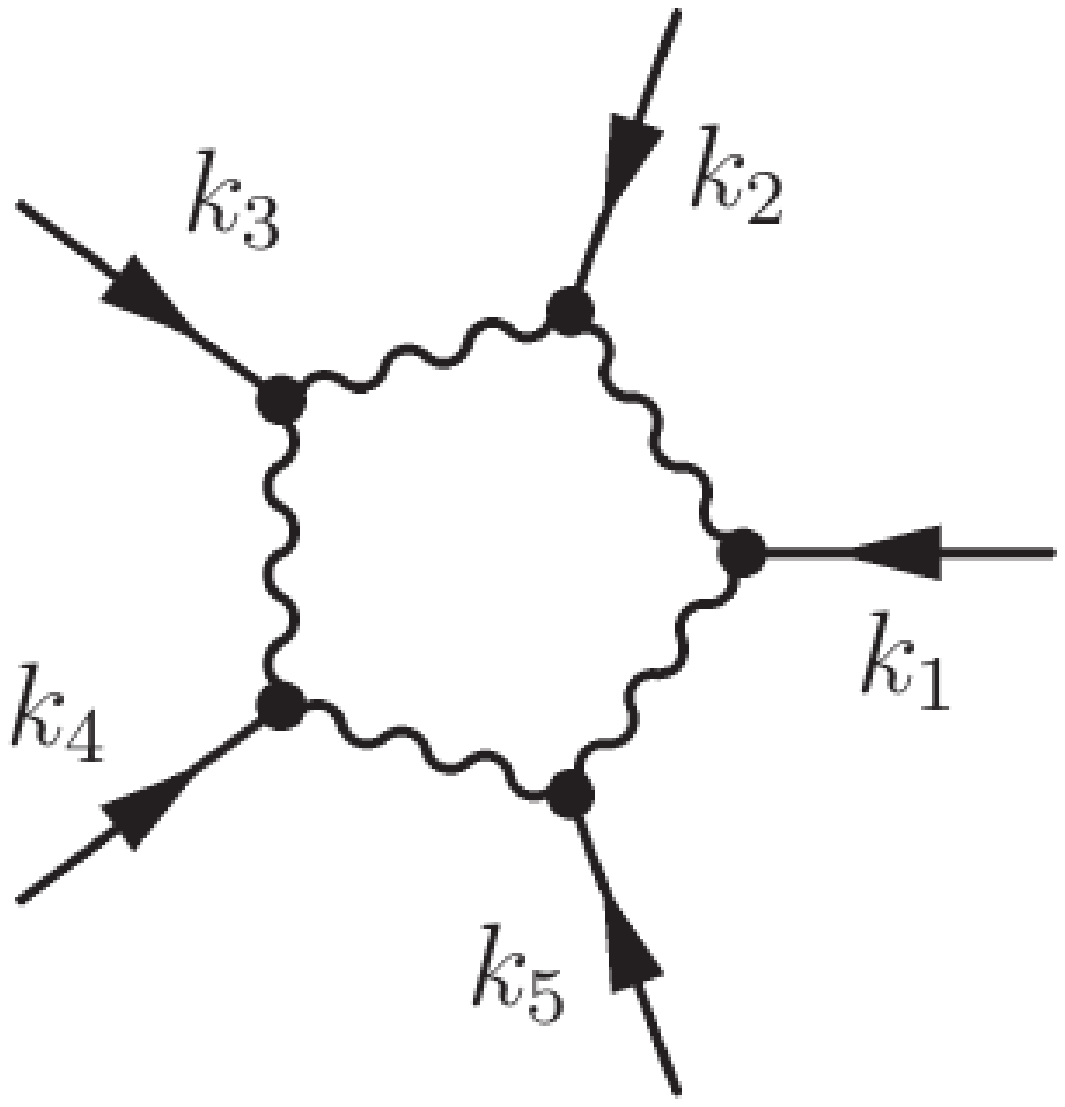}
\end{center}
\caption{Feynman diagram corresponding to the fifth cumulant where the solid and  wiggly lines represent the response and correlation
functions}
\end{figure}
\vskip2mm

We write Eq.\ \ref{h5c} in more compact form where the amputated part of the loop and the external legs are presented distinctly as 
\begin{eqnarray}
\langle h^5(\vx,t)\rangle_c=&& \int \frac{d^{d+1} \hat{k}_1}{(2\pi)^{d+1}}\int
\frac{d^{d+1} \hat{k}_2}{(2\pi)^{d+1}} \!\int \frac{d^{d+1}
\hat{k}_3}{(2\pi)^{d+1}}
\int \frac{d^{d+1} \hat{k}_4}{(2\pi)^{d+1}}
G(\hat{k}_1)  G(\hat{k}_2) G(\hat{k}_3)  \nonumber \\ 
&&  G(\hat{k}_4) G(-\hat{k}_1-\hat{k}_2-\hat{k}_3-\hat{k}_4)
L(\hat{k}_1,\hat{k}_2,\hat{k}_3,\hat{k}_4).
\label{W5total}
\end{eqnarray}
where $L$ is the amputated part of the loop diagram (excluding the external legs).
The unrenormalised (bare) form of the loop is written as 
\begin{eqnarray}
L^{(0)}(\hat{k}_1,\hat{k}_2, \hat{k}_3,\hat{k}_4) =&& 32
\left(\frac{-\lambda_0}{2}\right )^5 (2D_0)^5
\int \frac{d^{d+1}\hat{q_1}}{(2\pi)^{d+1}}
[\vqo \cdot(\vqo-\vko)][\vqo\cdot(\vqo+\vkt)]
\nonumber\\
&& \times [(\vqo+\vkt)\cdot(\vqo+\vkt+\vkth)]
[(\vqo+\vkt+\vkth)\cdot(\vqo+\vkt+\vkth+\vkf)] \nonumber \\
&& [(\vqf-\vkf)\cdot(\vko+\vkt+\vkth+\vkf+\vk_5)]
 G_0(\hat{q_1})G_0(\hat{k_1}-\hat{q_1}) \nonumber \\
&& \times G_0(-\hat{q}_1) G_0(\hat {k}_2+\hat{q}_1)
G_0(-\hat{q}_1-\hat{k}_2) G_0(\hat{k}_3+\hat{q}_1+\hat{k}_2)
G_0(-\hat{q}_1-\hat{k}_2-\hat{k}_3) \nonumber\\ 
 &&
G_0(\hat{k}_4+\hat{q}_1+\hat{k}_2+\hat{k}_3)G_0(-\hat{q}_1-\hat{k}_2-\hat{k}
_3-\hat{k}_4) G_0(\hat{q}_1+\hat{k}_5+\hat{k}_2+\hat{k}_3+\hat{k}_4) \nonumber\\ 
\label{L5initial}
\end{eqnarray} 
We shall find a renormalized equivalent of this bare quantity by means of a renormalization scheme 
employed earlier for the calculation of skewness and kurtosis \cite{Singha-Nandy14_PRE.90.062402,Singha-Nandy15_jstatmech}.
We perform the internal frequency integration in Eq.\ \ref{L5initial} and carry out the integration over the internal momentum  
restricted in the shell $\Lambda_0e^{-r} \leq q \leq \Lambda_0$, leading to 
\beq
L^{<}(r)=\frac{35}{8} \frac{\lambda^5_0 D^5_0 }{\pi \nu^9_0\Lambda^{7}_0}
\left[\frac{e^{7r}-1}{7}\right].
\label{L5r}
\eeq
Considering the iterative nature of the momentum shell elimination in the RG scheme, we obtain the differential equation 
\beq
\frac{dL}{dr}=\frac{35}{8} \frac{\lambda^5_0 D^5(r) }{\pi \nu^9(r)
\Lambda^{7}(r)}.
\label{diff_L_4}
\eeq
For large $r$, $\Lambda(r)=\Lambda_0 e^{-r}$ is identified as the  momentum  $k$.
Substituting the following functional forms 
\beq
\nu(k)=\lambda_0 \sqrt{\frac{D_0}{2\pi \nu_0}} k^{-1/2},
\label{nufk}
\eeq
and 
\begin{equation}
D(k)=\lambda_0 \sqrt{\frac{D_0}{2\pi \nu_0}} k^{-1/2},
\label{Dfk}
\end{equation}
we integrate the differential equation Eq. \ref{diff_L_4} and obtain 
\beq
L(r)= \frac{7\pi}{2} \lambda_0 \frac{D^3_0}{\nu^3_0 \Lambda^5_0}
e^{5r}.
\label{L5r}
\eeq
This expression for the renormalized loop needs to be expressed as a symmetric combination of the external momentum $\hat{k}_i$. 
Consequently $\Lambda_0 e^{-r}$ is expressed as the fully symmetric combination 
 \beq
L(\vko,0;\vkt,0;\vkth,0;\vkf,0)= \frac{7\pi}{2}\lambda_0
\left(\frac{D_0}{\nu_0}\right)^3 k^{-5/4}_1 k^{-5/4}_2 k^{-5/4}_3 k^{-5/4}_4
\label{L5k}
\eeq
for vanishing external frequencies. The frequency dependence is constructed by considering a scaling fuction of the form  
\beq
k_{i}^{5/4}f_5\left(\frac{\omega_{i}}{k_{i}^z}\right)=\frac{1}{k_{i}^{11/4}
\nu^2(k_{i})|G(k_{i},\omega_{i})|^2}.
\label{h5dsc}
\eeq
This form satisfies the property of real valuedness and it has the desired zero frequency (long time) limit. 
Consequently,  the renormalized loop in Fig. 1 is written as  
\begin{eqnarray}
L(\vko,\omega_1;\vkt,\omega_2;\vkth,\omega_3;\vkf,\omega_4)= && \frac{7\pi}{2} \lambda_0 
\left(\frac{D_0}{\nu_0}\right)^3 k^{11/4}_1 k^{11/4}_2 k^{11/4}_3
k^{11/4}_4 \nu^2(k_1) \nu^2(k_2)  \nu^2(k_3) \nu^2(k_4)  \nonumber \\
&& |G(k_1,\omega_1)|^2
|G(k_2,\omega_2)|^2 |G(k_3,\omega_3)|^2 |G(k_4,\omega_4)|^2. 
\label{L5dsc}
\end{eqnarray}
Substituting from Eq.\ \ref{L5dsc} in Eq.\ \ref{W5total}, and carrying out the frequency
integrations over $\omega_1$, $\omega_2$, $\omega_3$ and $\omega_4$, the fifth cumulant is obtained as 
\begin{eqnarray}
\langle h^5(\vx,t)\rangle_c= &&\frac{7}{4} \left(\frac{D_0}{2\pi
\nu_0}\right)^{5/2} \int^{\infty}_{-\infty}  dk_1 \int^{\infty}_{-\infty}
d k_2 \!\int^{\infty}_{-\infty} d k_3 \!\int^{\infty}_{-\infty} dk_4\!\nonumber
\\ 
&& J(k_1,k_2,k_3,k_4)
\label{h5afmomn}
\end{eqnarray}
where
\beq
J(k_1,k_2,k_3,k_4)=\frac{X(k_1,k_2,k_3,k_4)}{Y(k_1,k_2,k_3,k_4)}
\eeq
with
\begin{eqnarray}
X(k_1,k_2,k_3,k_4)=&&(-3|k_1|^6-3|k_2|^6-10|k_2|^{9/2}(2|k_3|^{3/2}
+2|k_4|^{3/2}+|k_1+k_2+k_3+k_4|^{3/2}) \nonumber \\
&&-10|k_1|^{9/2}(2|k_2|^{3/2}+2|k_3|^{3/2} +2|k_4|^{3/2}+|k_1+k_2+k_3+k_4|^{3/2}) \nonumber \\ 
&&-(|k_3|^{3/2}+|k_4|^{3/2}+ |k_1+k_2+k_3+k_4|^{3/2})^2 (3|k_3|^3+3|k_4|^3 + \nonumber \\ 
&& 4|k_4|^{3/2}|k_1+k_2+k_3+k_4|^{3/2} +|k_1+k_2+k_3+k_4|^3
\nonumber \\
&& +2|k_3|^{3/2}(7|k_4|^{3/2}+2|k_1+k_2+k_3+k_4|^{3/2})) \nonumber \\
&& 
-2|k_2|^3(17|k_3|^3+17|k_4|^3+23|k_4|^{3/2}|k_1+k_2+k_3+k_4|^{3/2}
 \nonumber \\ 
&& +6|k_1+k_2+k_3+k_4|^3
+|k_3|^{3/2}(62|k_4|^{3/2} +23|k_1+k_2+k_3+k_4|^{3/2}))
 \nonumber \\
&&-2|k_1|^3(17|k_2|^3 + 17|k_3|^3+17|k_4|^3 +23|k_4|^{3/2}|k_1+k_2+k_3+k_4|^{3/2}\nonumber \\
&&+6|k_1+k_2 +k_3+k_4|^3 +|k_3|^{3/2} (62|k_4|^{3/2}+23|k_1+k_2+k_3+k_4|^{3/2}) \nonumber \\
&&+ |k_2|^{3/2} (62|k_3|^{3/2}+62|k_4|^{3/2}+ 23 |k_1+k_2+k_3+k_4|^{3/2})))
\end{eqnarray}
and

\begin{eqnarray}
&Y(k_1,k_2,k_3,k_4)&= 256|k_1|^{5/4}|k_2|^{5/4}|k_3|^{5/4}|k_4|^{5/4} (|k_1|^{3/2}+|k-2|^{3/2}+|k_3|^{3/2} \nonumber \\
&&+|k_4|^{3/2}+|k_1+k_2+k_3+k_4|^{3/2})^5 \nonumber 
\end{eqnarray}
Considering the symmetry of the function $J(k_1,k_2,k_3,k_4)$, the integrations in Eq. \ref{h5afmomn} can be written as 
\begin{eqnarray}
\langle h^5(\vx,t)\rangle_c= &&\frac{7}{4} \left(\frac{D_0}{2\pi
\nu_0}\right)^{5/2} \int^{\infty}_{\mu}  dk_1 \int^{\infty}_{\mu}
d k_2 \!\int^{\infty}_{\mu} d k_3 \!\int^{\infty}_{\mu} dk_4 \nonumber \\
&& [2 J(k_1,k_2,k_3,k_4)+8 J(-k_1,k_2,k_3,k_4)\nonumber \\
&&+6 J(-k_1,-k_2,k_3,k_4)].
\label{h5inte}
\end{eqnarray}
where an infrared cut off $\mu$ has been set due to the infrared divergences in the integrations. 
We write the integrations as
\beq
J_1(\mu)= \int^{\infty}_{\mu}  dk_1 \int^{\infty}_{\mu}
d k_2 \!\int^{\infty}_{\mu} d k_3 \!\int^{\infty}_{\mu} dk_4
J(k_1,k_2,k_3,k_4),
\label{J1}
\eeq

\beq
J_2(\mu)= \int^{\infty}_{\mu}  dk_1 \int^{\infty}_{\mu}
d k_2 \!\int^{\infty}_{\mu} d k_3 \!\int^{\infty}_{\mu} dk_4
J(-k_1,k_2,k_3,k_4)
\label{J2}
\eeq
and 
\beq
J_3(\mu)= \int^{\infty}_{\mu}  dk_1 \int^{\infty}_{\mu}
d k_2 \!\int^{\infty}_{\mu} d k_3 \!\int^{\infty}_{\mu} dk_4
J(-k_1,-k_2,k_3,k_4)
\label{J3}
\eeq
So that Eq. \ref{h5inte} becomes
\beq
\langle h^5(\vx,t)\rangle_c= \frac{7}{4} \left(\frac{D_0}{2\pi
\nu_0}\right)^{5/2} [2 J_1(\mu)+8 J_2(\mu)+6 J_3(\mu)]
\label{h5-compact}
\eeq

The IR cut off dependent integrals are of the form   
\begin{eqnarray}
J_1(\mu)=a_1 \mu^{-5/2} \nonumber \\
J_2(\mu)=a_2 \mu^{-5/2} \nonumber \\
J_3(\mu)=a_3 \mu^{-5/2}
\label{J1J2J3}
\end{eqnarray}
where $a_1$, $a_2$ and $a_3$ are dimensionless constants.
Numerical integrations yield the dimensionless constants
as
\begin{eqnarray}
a_{1}=\lim_{\mu \rightarrow 0^+}[\mu^{5/2}J_{1}(\mu)]=0.00197 \nonumber \\
a_{2}=\lim_{\mu \rightarrow 0^+}[\mu^{5/2}J_{2}(\mu)]=0.00607  \nonumber \\
a_{3}=\lim_{\mu \rightarrow 0^+} [\mu^{5/2}J_{3}(\mu)]=0.00583
\label{dimcons}
\end{eqnarray}
via numerical convergences.
Substituting Eqs.\ \ref{dimcons}, \ref{J1}, \ref{J2}, \ref{J3}, \ref{J1J2J3} 
in Eq.\ \ref{h5-compact}, we finally obtain the expression for fifth cumulant as
\beq
\langle h^5(\vx,t)\rangle_c= \frac{7}{4}[2a_1+8a_2+6a_3]
\left(\frac{D_0}{2 \pi \nu_0}\right)^{5/2} \frac{1}{\mu^{5/2}}
\label{h5last}
\eeq
\section{Hyperskewness}
To calculate the hyperskewness, the expression for the second cumulant is needed.
 The second cumulant in the frequency and momentum space is expressed as
\begin{eqnarray}
\langle h^2(\vx,t)\rangle_c =\langle h^2(\vx,t)\rangle &=\int \frac{d^dk}{[2\pi]^{d}} \int \frac{d \omega}{[2\pi]} \int
\frac{d^dk'}{(2\pi)^{d}} \int \frac{d \omega'}{2\pi} \langle
 h(\vk,\omega)h(\vk',\omega')\rangle_c  \nonumber \\
& e ^{i[(\vk+\vk') \cdot \vx-(\omega+\omega')t]}
\label{h^2}
\end{eqnarray}
The Feynman loop corresponding to the correlation function is shown in Fig.\ 2. 
\begin{figure}[h!t]
\begin{center}
\includegraphics[width=4.6cm]{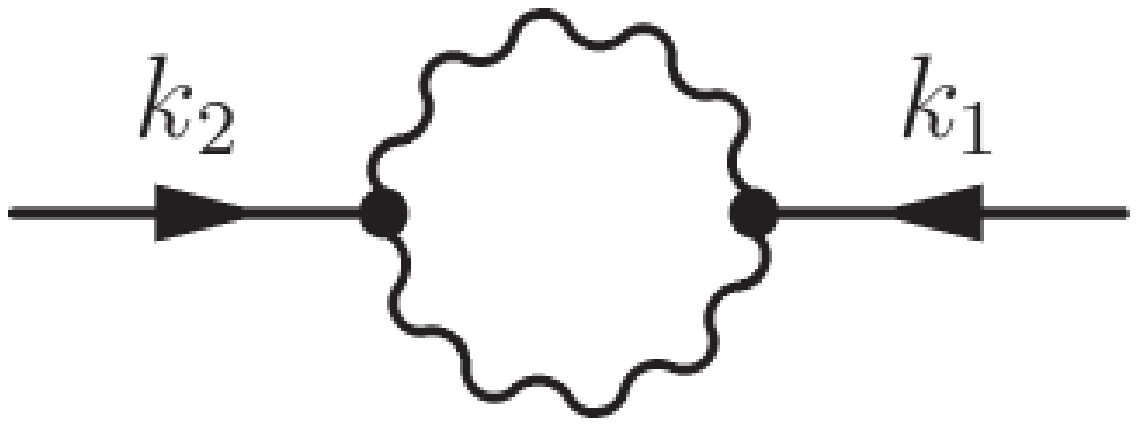}
\end{center}
\caption{Feynman diagram corresponding to the second cumulant where the solid and
wiggly lines represent the response and correlation functions }
\end{figure}

The frequency and momentum integrations in Eq. \ref{h^2} can be
evaluated \cite{Singha-Nandy14_PRE.90.062402,Singha-Nandy15_jstatmech} and it can be obtained as 
\beq
\langle h^2(\vx,t) \rangle_c=\frac{4}{\pi} \left(\frac{D_0}{2 \pi \nu_0}\right)
\frac{1}{\mu}
\label{eq=Q2}
\eeq
Using Eq.\  \ref {h5last} and \ref{eq=Q2} in \ref{defihypersk}, we obtain
\beq
\widetilde{S}=\frac{\langle h^5(\vx,t)\rangle_c}{\langle
h^2(\vx,t)\rangle_c^{5/2}}=\frac{7}{2}
\left(\frac{\pi}{4}\right)^{5/2}(a_1+4a_2+3a_3)
\label{hyper-b123}
\eeq
Substituting the values of $a_1$, $a_2$ and $a_3$ in Eq. \ref{hyper-b123} we obtain
\beq
\widetilde{S}=0.0835.
\label{hyprsk}
\eeq
\section{Discussion and Conclusion}
In this paper, we considered the stochastic growth of a surface on a flat geometry (in the stationary state) governed by the $(1+1)$-dimensional
KPZ equation. We followed a diagrammatic RG scheme to evaluate the Feynman diagram [Fig.\ 1]  for the fifth cumulant at one loop order 
corresponding to the $(1+1)$-dimensional KPZ dynamics. We started with the unrenormalized loop with bare  parameters ($\nu_0$, $\lambda_0$ and $D_0$ ) 
and invoked a shell elimination scheme (belonging to the shell $\Lambda_0 e^{-r}\leq q \leq \Lambda_0$) to obtain a differential equation 
(Eq. \ref{diff_L_4}) representing the recursion relation for successive elimination of momenta in thin shells. 
The solution of the differential equation yielded the renormalized expression (Eq. \ref{L5r}) for the loop diagram. This facilitated the evaluation 
of the cumulant $\langle h^5(\mb x, t)\rangle_c$ given by Eq. \ref{W5total} involving renormalized quantities.  The resulting integrals are found
to be infrared divergent and hence  $\langle h^5(\mb x, t)\rangle_c$ depends on the infrared cutoff $\mu$. The momentum integrals are
evaluated numerically to obtain  $\langle h^5(\mb x, t)\rangle_c$ as given by  Eq. \ref{h5last}. Normalizing this value with respect to the  
$\langle h^2(\mb x, t)\rangle^{5/2}_c$ resulted in the hyperskewness value as $\widetilde{S}=0.0835$. It is interesting to note that all parameters
of the KPZ dynamics ($\nu_0$, $\lambda_0$, $D_0$) and the momentum cutoffs ($\Lambda_0$ and $\mu$) finally cancel out to yield this value, suggesting 
its universality. 

Although there have been many studies on the lower order normalized moments such as skewness and kurtosis, the study of hyperskewness
remains a rarity. However, there have been studies based on numerical models that focus on the probability distribution function 
belonging to the KPZ universality class \cite{Michael_Herbert_00,Takeuchi_prl_110_2013,Healy_Lin_PRE.89.010103}. In the steady state, 
this distribution has been shown to be identical with the Baik-Rains distribution \cite{Takeuchi_prl_110_2013,Healy_Lin_PRE.89.010103}  
with zero mean \cite{Baik-Rains-JSP-160, Michael_Herbert_00}. 

The universal Baik-Rains $F_0$ distribution is a function of the solutions of Painlev{\'e}-II equation, 
namely, $u''(x)=2 u^3(x)+x u(x)$. Hastings and McLeod \cite{Hastings-McLeod_1980} obtained a unique solution with 
the asymptotic boundary conditions $u(x) \sim -Ai(x)$ as $x \rightarrow \infty$ and $u(x) \sim -\sqrt{-x/2}$ as 
$x \rightarrow -\infty$ where $Ai(x)$ is the Airy function. To solve the Painlev{\'e}-II equation, 
Tracy and Widom performed a numerical integration \cite{Tracy-Widom-1994}. Subsequently, 
Pr{\"a}hofer and Spohn \cite{Prahofer-Spohn_JSP_2004} obtained an arbitrary 
high precession solution by performing Taylor expansions. Baik-Rains distribution is obtained from these solutions via known 
mathematical relations. On the other hand, the Fredholm determinant representation of the random matrix theory has been observed 
to be  conceptually simpler and numerically efficient than Painlev{\'e}-II to obtain the probability 
distributions \cite{Bornemann_arxiv_0904.1581} ($F_2$, $F_1$, $F_0$, etc.).

It may however be noted that the Baik-Rains value for hyperskewness is $0.3092$ 
which may be obtained from Fredholm determinant representation\cite{Bornemann_arxiv_0904.1581,Bornemann-private} as well as from the 
numerical data \cite{Prahoferdata} via solution of the Painlev{\'e}-II. This Baik-Rains value is distinctly higher than our calculated value 
$\tilde S=0.0835$. This underestimation via the RG calculation is due to the fact that the method is based on a perturbative approach where 
the calculation is carried out only at one-loop order. Since the dynamics is governed by a white noise following a Gaussian distribution, 
the lowest order approximation appeares to be  influenced rather strongly by the Gaussian noise. A similar trend was observed in the one-loop 
perturbative calculation for kurtosis, namely, $Q=0.1523$ \cite{Singha-Nandy15_jstatmech} which is lower than the Baik-Rains value $0.2892$. 
However the value for skewness via the perturbative scheme, namely, $S=0.3237$ \cite{Singha-Nandy14_PRE.90.062402} is slightly lower than the 
Baik-Rains value $0.3594$. Thus it appears that the Gaussian white noise plays a more dominant role in the perturbative calculations for the
higher order moments, namely, kurtosis and hyperskewness. More involved calculations with the incorporation of higher order contributions in 
the perturbative expansion are expected to yield better estimates for these higher order moments.

We note that it is next to impossible to obtain a closed-form analytical expression for the full probability distribution 
starting with the KPZ equation driven by the stochastic noise. Consequently, in an analytical approach, one usually evaluates a few 
lower and higher order moments (such as skewness, kurtosis and hyperskewness) from the governing dynamics. This outlook has motivated
us to evaluate these numbers directly from the $(1+1)$-dimensional KPZ dynamics in the stationary state. 

\subsection*{Acknowledgements} 
T.S. is thankful to the Ministry of Human Resource Development (MHRD), Government of India, 
for financial support through a scholarship. M.K.N. is indebted to the Indian Institute of 
Technology Delhi, and particularly to Prof. Senthilkumaran and Prof. Ravishankar, Department
of Physics, for hospitality and for extending various facilities at IIT Delhi.

\end{document}